\documentclass[aps,prl,twocolumn,floatfix,amsfonts]{revtex4-1}
\usepackage{tabularx}
\usepackage{graphicx}
\usepackage{amsmath,amsthm,amssymb}
\pdfpageattr {/Group << /S /Transparency /I true /CS /DeviceRGB>>}

\begin{document}

\title{Kondo effect in three-dimensional Dirac and Weyl systems}

\author{Andrew K. Mitchell}
\author{Lars Fritz}
\affiliation{Institute for Theoretical Physics, Utrecht University, 3584 CE Utrecht, The Netherlands}


\begin{abstract}
\noindent Magnetic impurities in three-dimensional Dirac and Weyl systems are shown to exhibit a fascinatingly diverse range of Kondo physics, with distinctive experimental spectroscopic signatures. When the Fermi level is precisely at the Dirac point, Dirac semimetals are in fact unlikely candidates for a Kondo effect due to the pseudogapped density of states. However, the influence of a nearby quantum critical point leads to the unconventional evolution of Kondo physics for even tiny deviations in the chemical potential. Separating the degenerate Dirac nodes produces a Weyl phase: time-reversal symmetry-breaking precludes Kondo due to an effective impurity magnetic field, but different Kondo variants are accessible in time-reversal \emph{invariant} Weyl systems. 
\end{abstract}
\maketitle


The last decade has seen enormous activity in the research of electronic topological states of matter, sparked by the discovery of topological insulators (TIs), protected by time-reversal symmetry (TRS)  \cite{kane2005z,*kane2005quantum,bernevig2006quantum,moore2007topological,fu2007topological,konig2007quantum,xia2009observation,hasan2010colloquium,qi2011topological}. More recently it was appreciated that \emph{gapless bulk states} can also be topologically nontrivial~\cite{wan2011topological,yang2011quantum,burkov2011weyl,halasz2012time,xu2011chern,young2012dirac}. Experimental candidates for such three-dimensional (3D) Dirac semimetals (SMs), such as Na$_3$Bi and Cd$_3$As$_2$, are now emerging~\cite{liu2014discovery,liu2014stable,borisenko2014experimental,jeon2014landau,neupane2014observation,he2014quantum}. 

Dirac SMs possess degenerate Dirac nodes at isolated points of the Brillouin zone, around which electronic excitations behave as 3D massless Dirac fermions. 
As such, Dirac SMs are 3D counterparts of graphene~\cite{liu2014discovery}, whose doubly-degenerate Dirac cones are protected by TRS and lattice inversion symmetry (IS). 

The degeneracy can be lifted by perturbations breaking either TRS or IS. Nodes of opposite chirality are then separated in momentum space or energy in the resulting topological Weyl phase. This separation produces remarkable stability to interactions \cite{burkov2011weyl}, and unique physical properties such as the `chiral anomaly' \cite{zyuzin2012topological} and surface Fermi arcs \cite{wan2011topological}. Several Weyl SM materials have recently been discovered \cite{weylgen1,weylgen2,weylgen3} -- including TR-invariant systems such as TaAs and NbAs \cite{weylexpt1,weylexpt2,weylexpt3}.

\begin{figure*}[t]
\includegraphics[width=0.9\textwidth]{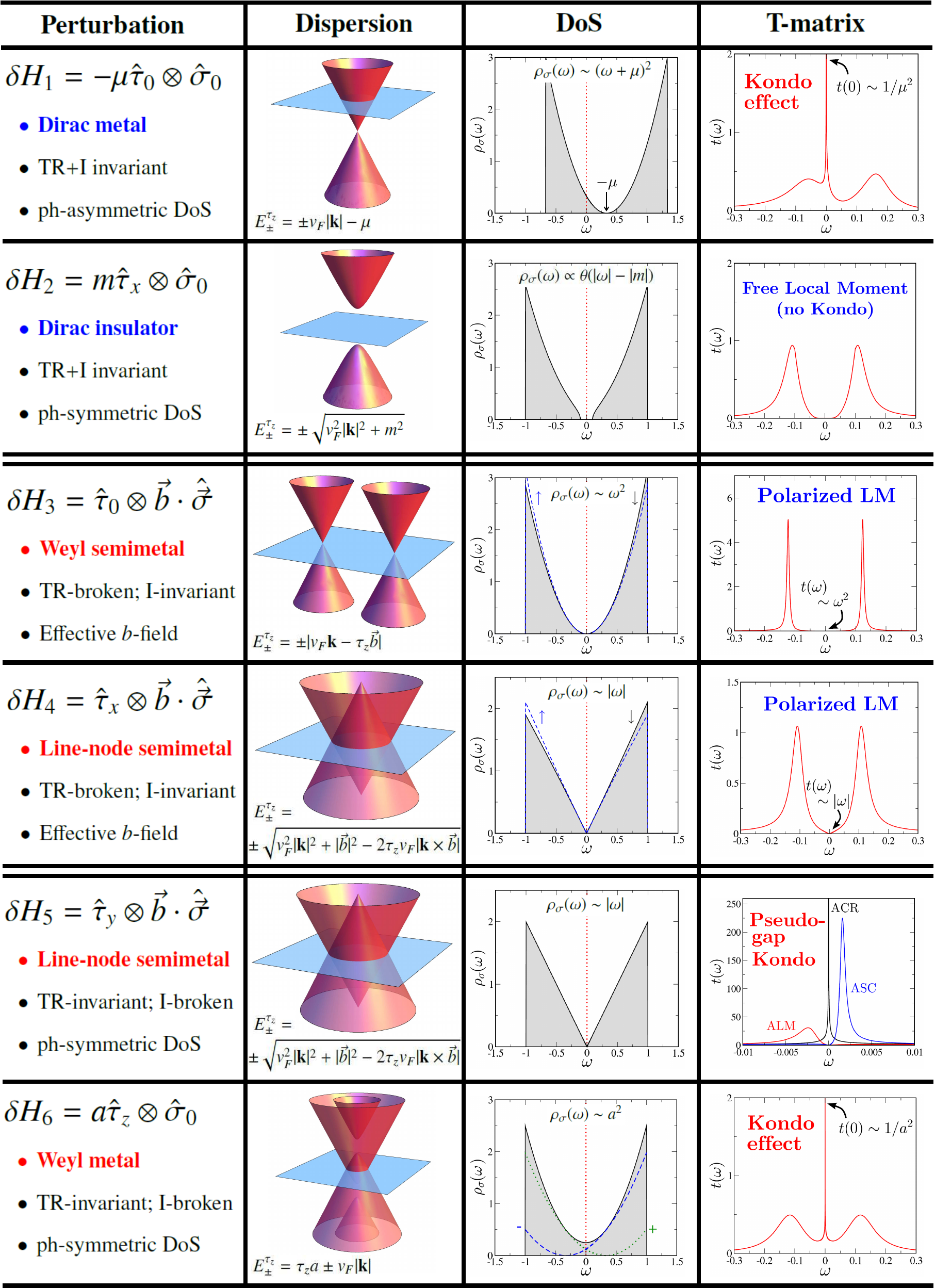}
\caption{
Summary of possible perturbations to the Dirac semimetal, $\delta H_1$ to $\delta H_6$ (top to bottom), and their effect on the dispersion (sketched in the space of $k_x$, $k_y$ and $E$ for fixed $k_z$), spin-resolved density of states $\rho_{\sigma}(\omega)$, and scattering T-matrix spectrum $t(\omega)$ at $T=0$ (left to right). Shown for band cutoff $D=v_{F}\equiv 1$ and magnetic impurity interaction strength $U=0.3$. 
Impurity asymmetry $\eta=-2\epsilon/U=1$ and impurity-host coupling $g_{\tau}\equiv g=0.2$ for $\delta H_{1,2,3,4,6}$, while for $H_{5}$ we use $\eta=\tfrac{1}{2}$ and $g=0.42, 0.479, 0.52$ for the ALM, critical ACR, and ASC results, respectively. Perturbation strength $\mu=-\tfrac{1}{3}$ for $\delta H_1$, $m=0.1$ for $\delta H_2$, $b_z=0.1$ for $\delta H_{3,4,5}$, and $a=\tfrac{1}{3}$ for $\delta H_6$. The T-matrix results in the fourth column are calculated for the impurity-coupled system using the DoS shown in the third column, obtained exactly by NRG.
}\label{fig:summary}
\end{figure*}

\begin{figure}[b]
\includegraphics[width=70mm]{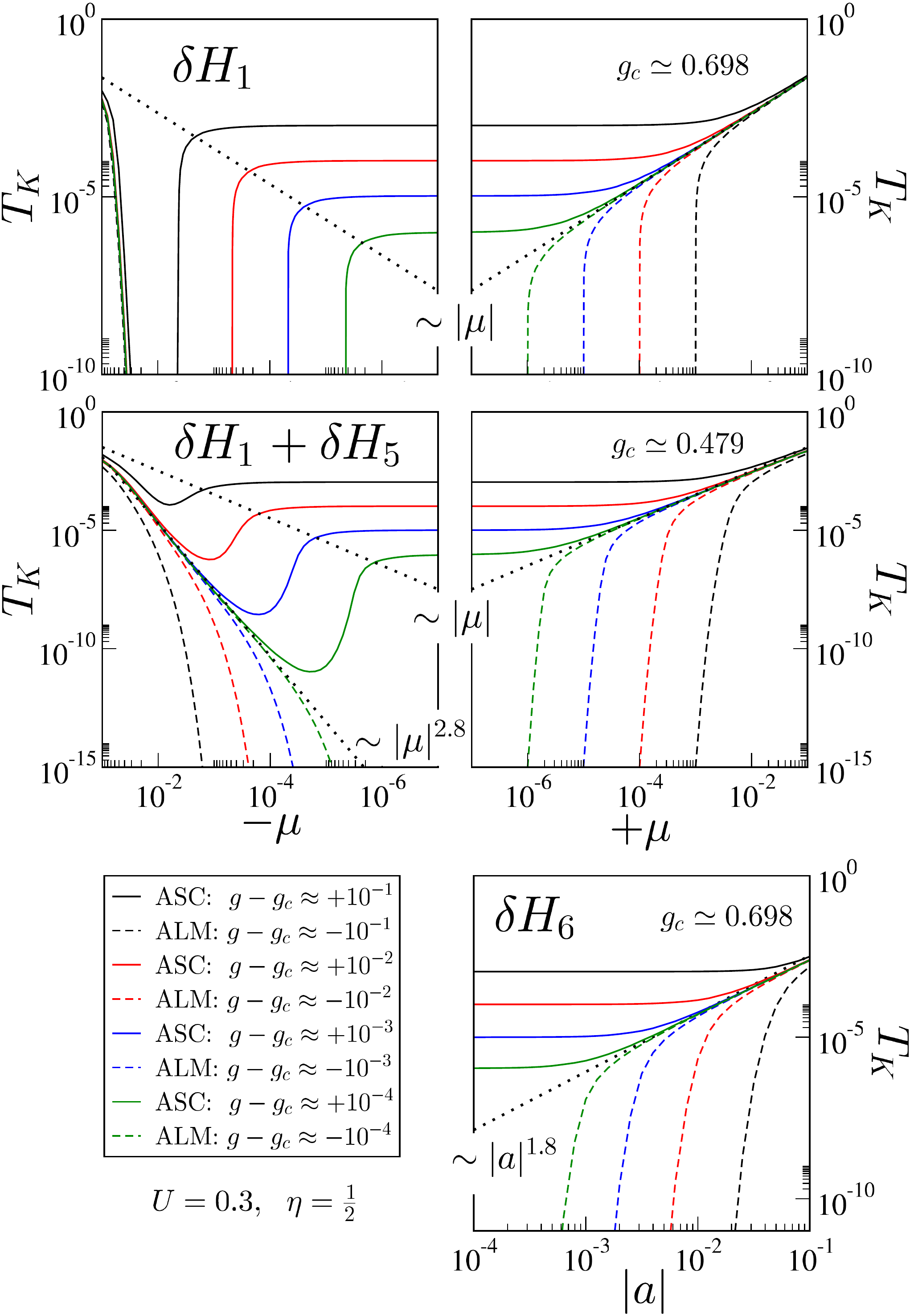}
\caption{Evolution of the Kondo temperature in the three situations where the Kondo effect can manifest. $T_K$ vs chemical potential $\mu$ shown for the Dirac metal with $\delta H_1$, and the  line-node Weyl semimetal with $\delta H_1+\delta H_5$, in the top and middle panels, respectively. Bottom panel shows $T_K$ vs $a$ in the Weyl metal with $\delta H_6$. Dotted lines show the powerlaw crossovers. 
}\label{fig:tk}
\end{figure}

In this paper we discuss Dirac and Weyl systems with isolated \emph{magnetic impurities} (e.g.\ transition metal adatoms), the scattering from which producing a diverse range of Kondo physics whose characteristic experimental signatures fingerprint the topological host. We show that the host density of states (DoS), and the symmetries broken by perturbations, play a key role in determining the kind of Kondo effect that can take place \cite{note:preprints}. Perturbations can lead to Dirac metals and insulators or various Weyl variants; in each case, different Kondo physics manifest, ranging from standard metallic, through quantum critical, to pseudogap Kondo. Fig.~\ref{fig:summary} summarizes these findings. 
In all Kondo phases, we uncover an unconventional evolution of the Kondo temperature when the Fermi level is tuned in the vicinity of the Dirac point --- see Fig.~\ref{fig:tk}. 

The theoretical framework we develop applies to impurities in Dirac materials such as Na$_3$Bi and Cd$_3$As$_2$; we show that strong particle-hole asymmetry (which can be on either the impurity or in the high-energy part of the host DoS) can stabilize a pseudogapped Kondo state. On doping the Fermi level away from the Dirac point, the Kondo effect ubiquitously appears, although influence of a nearby quantum critical point is important.
In Weyl systems, we show that the specific way in which the nodes are separated in momentum space is important. When it is achieved by TRS breaking (e.g.\ due to magnetic ordering, as proposed in pyrochlore iridates \cite{wan2011topological}), the Kondo effect is suppressed. When the nodes are separated due to IS breaking (e.g.\ in the newly-discovered TRS-invariant but non-centrosymmetric Weyl materials TaAs and NbAs \cite{weylexpt1,weylexpt2,weylexpt3}), pseudogap Kondo can again be realized. Any given microscopic Dirac/Weyl realization can be classified according to its symmetries, and can be compared to one of the generic situations discussed in this Letter. Alternatively, details of the microscopic structure can be inferred from the type of Kondo effect observed.


{\it {Bulk model and symmetries:}}
The clean, unperturbed Dirac SM is described by the minimal Bloch Hamiltonian, 
\begin{eqnarray}\label{eq:DiracSM}
{\hat{\mathcal{H}}}_{\rm{D}} ({\bf{k}})=v_F \; {\hat{\tau}}_z  \otimes {\bf{k}}\cdot {\hat{{\vec{\sigma}}}}\;,
\end{eqnarray}
where $\hat{\sigma}$ and $\hat{\tau}$ are Pauli matrices acting respectively in spin and orbital space, and $v_F$ is the effective Fermi velocity. 
The pristine Dirac SM possesses both TRS and IS --- meaning  
$\mathcal{T}{\hat{\mathcal{H}}}_{\rm{D}}({\bf{k}})\mathcal{T}^{-1}={\hat{\mathcal{H}}}_{\rm{D}}(-{\bf{k}})$ in terms of the time-reversal operator $\mathcal{T}=\hat{\tau}_0 \otimes (i \hat{\sigma}_y)K$ (with complex conjugation $K$); and $\mathcal{P}{\hat{\mathcal{H}}}_{\rm{D}}({\bf{k}})\mathcal{P}^{-1}={\hat{\mathcal{H}}}_{\rm{D}}(-{\bf{k}})$, with inversion operator $\mathcal{P}=\hat{\tau}_x \otimes \hat{\sigma}_0$.


{\it{Bulk perturbations:}}
The description of real systems necessitates inclusion of perturbations to Eq.~\eqref{eq:DiracSM}; in particular, the node-separation in Weyl systems is by definition a result of these perturbations, and therefore they must be considered explicitly. 
To leading order, perturbations are of form $\delta\hat{H}=\left ( \vec{a} \cdot \hat{\vec{\tau}} +a_0\hat{\tau}_0 \right )\otimes  \left (\vec{b}\cdot \hat{\vec{\sigma}} + b_0 \hat{\sigma}_0 \right)$. Microscopic models of Dirac/Weyl systems map onto ${\hat{\mathcal{H}}}_{\rm{D}} ({\bf{k}})+\delta\hat{H}$ at low energies. Depending on the parameters $\vec{a}$, $a_0$, $\vec{b}$, and $b_0$, the perturbations can leave the Dirac system TRS and IS invariant, break either TRS or IS to produce a Weyl system, or break both TRS and IS \cite{note:perts}. Representative examples are discussed below; see also Fig.~\ref{fig:summary}.


{\it{Impurity model:}} 
We model the magnetic impurity 
as a single correlated quantum level, coupled locally in real-space to conduction electrons of the host Dirac or Weyl system. The Andersonian Hamiltonian \cite{hewson1997kondo} is $H=H_{\text{host}}+H_{\text{imp}}+H_{\text{hyb}}$, where 
$H_{\rm{host}}= \int \frac{d^3 k}{(2 \pi)^3} \Psi^\dagger ({\bf{k}}) [\hat{\mathcal{H}}_{\text{D}}(\textbf{k})+\delta\hat{H}]\Psi ({\bf{k}})$ in terms of conduction electron operators $\Psi ({\bf{k}})$ in $\tau$- and $\sigma$-space. The impurity is described by 
$H_{\rm{imp}}=\sum_{\sigma} \epsilon d^\dagger_{\sigma} d^{\phantom{\dagger}}_{\sigma}+U d^\dagger_{\uparrow} d^{\phantom{\dagger}}_{\uparrow} d^\dagger_{\downarrow} d^{\phantom{\dagger}}_{\downarrow}$, with $d_\sigma$ a spin $\sigma=\uparrow,\downarrow$ operator for the impurity level. $\eta=-2\epsilon/U$, characterizes the impurity particle-hole (PH) asymmetry. The impurity-host coupling is specified by 
$H_{\rm{hyb}}= \sum_{\sigma,\tau} g^{\phantom{\dagger}}_{\sigma\tau} d^{\dagger}_{\sigma}\Psi^{\phantom{\dagger}}_{\sigma,\tau}({\bf{x}}=0)+{\rm{H.c.}}$. 
The exact retarded impurity Green function, after integrating out bulk fermions, reads ${\hat{G}}^{\text{imp}}(z)=[ (z-\epsilon) {\hat{{\mathbb{I}}}}-\hat{\Delta}(z) - \hat{\Sigma}(z)]^{-1}$, where the hybridization matrix is given by $\hat{\Delta}_{\sigma,\sigma'}(z)=\sum_{\tau,\tau'} g^{\phantom{*}}_{\sigma\tau} g_{\sigma'\tau'}^{*} {\hat{G}}_{\sigma\tau,\sigma'\tau'}^{\rm{host}}(z) $ in terms of \emph{local} host Green functions at the impurity position ${\bf{x}}=0$, given by ${\hat{G}}^{\rm{host}}(z)=\int \frac{d^3 k}{(2 \pi)^3} [(z-\epsilon) {\hat{{\mathbb{I}}}}-\hat{\mathcal{H}}_{\text{D}}(\textbf{k})-\delta\hat{H}]^{-1}$. 
Depending on the microscopic details of how the impurity is embedded in the Dirac/Weyl host material, the coupling $g_{\sigma\tau}$ may depend explicitly on $\sigma$ and $\tau$. However, the hybridization is a $2\times 2$ matrix in spin space, and involves a trace over the pseudospin degree of freedom $\tau$. Ultimately, the Kondo physics is controlled by the symmetries of $\hat{\Delta}(z)$. Although for concreteness we now set $g_{\sigma\tau}\equiv g$, real systems should be comparable to one of the distinct cases discussed here. 

$\hat{\Sigma}(z)$ is the interaction self-energy, which contains nontrivial correlation effects due to the impurity. We obtain it exactly from the numerical renormalization group (NRG) \cite{bulla2008numerical,note:nrg}, whose input is $\hat{\Delta}(z)$. 


{\it{T-matrix:}} 
The T-matrix describes scattering in the Dirac or Weyl system due to the impurity, ${\hat{\mathcal{G}}}^{\rm{host}}(z)={\hat{G}}^{\rm{host}}(z)[{\hat{{\mathbb{I}}}}+\hat{T}(z){\hat{G}}^{\rm{host}}(z)]$, with elements $\hat{T}_{\sigma\tau,\sigma'\tau'}(z)=g^{*}_{\tau}g_{\tau'}^{\phantom{*}}{\hat{G}}^{\text{imp}}_{\sigma,\sigma'}(z)$. Related to the impurity DoS, the T-matrix can be probed locally by scanning tunneling spectroscopy. Quasiparticle interference is also sensitive to the T-matrix \cite{mitchell2013kondo,derry2015quasiparticle,*mitchell2015multiple}, as is resistivity \cite{costi1994transport}. Here, we consider its spectrum, $t(\omega)=-\tfrac{1}{\pi}\text{Im Tr}~\hat{T}(\omega+i0^+)$.


{\it{Kondo physics in Dirac systems:}} 
The energy eigenvalues of $\hat{\mathcal{H}}_{\text{D}}(\textbf{k})$ are two-fold degenerate. The local host DoS in 3D, $\rho_{\sigma}(\omega)=-\tfrac{1}{\pi}\text{Im}\sum_{\tau}{\hat{G}}^{\rm{host}}_{\sigma\tau,\sigma\tau}(\omega+i0^{+})\propto \omega^2$, is therefore \emph{quadratic} at low energies, since the Dirac point is at the Fermi level.  Furthermore, $\hat{\Delta}(z)$ is diagonal in spin-space, and so the effective impurity problem falls into the pseudogapped Kondo class \cite{gonzalez1998renormalization,fritz2004phase,*vojta2004upper,logan2014common}, with intact TRS. With $\eta=1$ on the impurity, the entire system is PH symmetric; the depleted DoS strictly precludes the possibility of a Kondo effect. Below temperature/energy scales $\sim U$, a spin-$\tfrac{1}{2}$ local moment (LM) forms which remains unscreened down to $T=0$.

When PH symmetry is broken, $\eta\ne 1$, the impurity can be Kondo screened --- there is a quantum phase transition from the asymmetric LM (ALM) phase to an asymmetric strong coupling (ASC) Kondo phase, on increasing the impurity-host coupling $g$, with a nontrivial interacting critical point (ACR) at $g_c$ \cite{fritz2004phase,*vojta2004upper}. The Kondo temperature in the ASC phase vanishes as $T_K\sim |g-g_c|$ on approaching the critical point $g\rightarrow g_c^{+}$. Incipient RG flow in the vicinity of ACR arises in both ASC or ALM phases for small $|g-g_c|\ll U$. We have confirmed this scenario for the Dirac SM using NRG, albeit that a rather large coupling $g$ is required to access ASC. 

The effects of perturbing the pristine Dirac SM are discussed below in relation to Fig.~\ref{fig:summary} (top to bottom for different perturbation types), showing in particular the exact T-matrix in the right column. 
In Fig.~\ref{fig:tk} we study the unconventional evolution of the Kondo temperature $T_K$, in all cases where the Kondo effect manifests. 


(I) $\delta H_1=-\mu \hat{\tau}_0 \otimes \hat{\sigma}_0$ corresponds to a chemical potential and does not break TRS or IS --- see first row of Fig.~\ref{fig:summary}. The low-energy DoS $\rho_{\sigma} (\omega) \propto (\omega+\mu)^2$ is finite at the Fermi level, implying that the impurity spin is screened on the lowest energy scales. For large $\mu$, we indeed find a finite $T_K$, which scales in the `standard' way for magnetic impurities in metals \cite{hewson1997kondo}, 
\begin{eqnarray}\label{eq:standardtk}
T_K\sim e^{-1/(\rho_{\sigma}(0)J)} \;\; ; \;\;J \propto g^2[U\eta(2-\eta)]^{-1} \;,
\end{eqnarray}
with $J$ the effective Kondo exchange. The top-right panel of Fig.~\ref{fig:summary} shows the spectrum of the T-matrix due to impurity scattering at $T=0$, exhibiting the classic three-peak structure. The central Kondo resonance (of width proportional to $T_K$) embodies enhanced spin-flip scattering at low energies, and is pinned at the Fermi level $\propto 1/\mu^2$ by the Friedel sum rule \cite{hewson1997kondo}. Note the pronounced PH asymmetry of the spectrum, due to the asymmetric DoS.

A metallic Kondo effect would imply the scaling $\ln T_K \sim -1/\mu^2$, meaning the Kondo temperature becomes very small (and effectively unobservable in experiment) when the Fermi level is near the Dirac point. However, this analysis breaks down at small $\mu$ due to the nearby ACR quantum critical point. 

For an asymmetric impurity $\eta < 1$, the impurity is Kondo-screened in the ASC phase for $g>g_c$, even at $\mu=0$. The Kondo temperature in fact evolves smoothly from this finite $\mu=0$ value to Eq.~\ref{eq:standardtk} on increasing $\mu>0$. This is shown in the top-right panel of Fig.~\ref{fig:tk} (solid lines). By contrast, $T_K=0$ in the ALM phase for $g<g_c$ at $\mu=0$, implying $T_K\rightarrow 0$ as $\mu\rightarrow 0$ (see dashed lines). The influence of the critical point therefore extends well into the `standard' Kondo phase. Indeed, the finite-$\mu>0$ Kondo phase is separated into a `high-$T_K$ regime' and a `low-$T_K$ regime' along $T_K\sim |g-g_c|$ (dotted lines, Fig.~\ref{fig:tk}), depending on the parent $\mu=0$ phase.

PH symmetry is broken by both $\eta\ne 1$ and $\mu\ne 0$, which can either reinforce or partially cancel. Varying $\mu$ therefore changes the effective proximity to the critical point. For a given $\eta$, this leads to a pronounced asymmetry between $\mu>0$ and $\mu<0$, as seen in the top panels in Fig.~\ref{fig:tk}. Indeed, \emph{increasing} $|\mu|$ can drastically reduce $T_K$ by tipping the system from the high-$T_K$ (ASC parent) regime to the low-$T_K$ (ALM parent) regime. Only for very large $|\mu|$ is Eq.~\ref{eq:standardtk} recovered.


 (II) $\delta H_2 = m \hat{\tau}_x \otimes \hat{\sigma}_0$ opens a gap --- see Fig.~\ref{fig:summary} (second row). There is no Kondo screening at PH symmetry because there are no low-energy electronic degrees of freedom to build up the Kondo singlet. The ground state is therefore a free LM. Although the T-matrix for this system at $T=0$ contains Hubbard satellites at energies $\epsilon$ and $(U+\epsilon)$ due to charge fluctuations, it is fully gapped at low energies, as verified by NRG. For PH asymmetric systems, it is known from semiconductors/insulators \cite{chen1998kondo} that a first-order transition between LM and Kondo singlet states arises on increasing $g$. In principle, mid-gap poles in the T-matrix could appear. But in the present context of impurities in Dirac insulators, this scenario seems unlikely, due to the combination of strong PH asymmetry and large $g$ which we find are required.


{\it{TRS-broken Weyl systems:}} 

(III) $\delta H_3= \hat{\tau}_0 \otimes \vec{b}\cdot \hat{\vec{\sigma}}$ splits the two degenerate Dirac theories in momentum space --- see Fig.~\ref{fig:summary} (third row). Without loss of generality, we choose $\vec{b}\equiv b_z$, such that $\hat{G}^{\text{imp}}$ is diagonal. 
Although the low-energy DoS for $\sigma=\uparrow$ and $\downarrow$ has the same quadratic form to leading order, the bulk TRS-breaking leads to an \emph{effective magnetic field} on the impurity, $\hat{\Delta}_{\uparrow\uparrow}(0)-\hat{\Delta}_{\downarrow\downarrow}(0)\propto b_z$. The impurity is fully polarized on the temperature/energy scale of $b_z$, and no Kondo effect can occur. This is manifest in the (spin-summed) T-matrix as sharp peaks, each of weight $\simeq 4g^2$. At low energies, the spectrum must decay with the same power as the hybridization, $t(\omega)\sim \omega^2$.


(IV) $\delta H_4= \hat{\tau}_x \otimes \vec{b}\cdot \hat{\vec{\sigma}}$: one of the two Dirac theories is gapped out, while the other theory acquires a line node --- see Fig.~\ref{fig:summary} (fourth row). This line-node Weyl SM has a \emph{linear} DoS \cite{note:linenode}. TRS-breaking induces a finite effective impurity field, and Kondo is again quenched. The low-energy behavior of the T-matrix is now linear, $t(\omega)\sim |\omega|$.


{\it{TRS-invariant Weyl systems:}} 

(V) $\delta H_5= \hat{\tau}_y \otimes \vec{b}\cdot \hat{\vec{\sigma}}$ realizes a line-node Weyl SM as for case (IV), but without an impurity field, $\hat{\Delta}_{\uparrow\uparrow}(z)=\hat{\Delta}_{\downarrow\downarrow}(z)$. The low-energy DoS is linear \cite{note:linenode}, $\rho_{\sigma}(\omega) \propto |\omega|$. A magnetic impurity in this line-node SM is therefore a good candidate to realize pseudogap Kondo physics: it is more likely accessible than in the Dirac SM case because the DoS vanishes less rapidly near the Fermi level. The ASC phase can therefore be reached with moderate impurity PH asymmetry $\eta$ and smaller impurity-host coupling $g$. The fifth row of Fig.~\ref{fig:summary} contains NRG results for the T-matrix in the ALM/ASC phases, and at the ACR critical point, each showing distinctive signatures. 

As with case (I), the pseudogap Kondo critical point has a significant influence on the `regular' Kondo phase at finite $\mu$, demonstrated by the unconventional evolution of $T_K$ in Fig.~\ref{fig:tk} (middle panels).
For $\eta<1$ and $\mu>0$, there is again a separatrix between high- and low-$T_K$ regimes following $T_K\sim |\mu|$. At large $\mu$, the Kondo temperature varies more slowly than in the Dirac SM, $\ln T_K \sim -1/|\mu|$. However, the competition between asymmetry coming from the impurity ($\eta\ne 1$) and host ($\mu\ne 0$) is more finely balanced due to the linear DoS. This is starkly evident for $\mu<0$, where a nontrivial intermediate regime of two-stage Kondo screening is found between $T_K\sim |\mu|$ and $T_K\sim |\mu|^{\xi}$ (dotted lines), with $\xi\simeq 2.8$ extracted from NRG. In this regard, the TRS-invariant line-node SM is similar to graphene, where equivalent impurity effects have previously been found \cite{vojta2010gate,*fritz2013physics}.


(VI) $\delta H_6= a \hat{\tau}_z \otimes \hat{\sigma}_0$ splits the two degenerate Dirac theories in energy --- see sixth row, Fig.~\ref{fig:summary}. The DoS is nonvanishing everywhere, and one expects regular Kondo physics (see e.g.~the spectrum of the T-matrix, plotted for large perturbation strength $a=\tfrac{1}{3}$). For small $a$, we again see the influence of the parent $a=0$ pseudogap Kondo quantum critical point, as shown in Fig.~\ref{fig:tk} (bottom panel). The situation here is distinct from cases (I) or (V), since the perturbation $\delta H_6$ does not introduce additional PH breaking. We find from NRG a new and nontrivial powerlaw separatrix between high- and low-$T_K$ regimes, $T_K\sim |a|^{\zeta}$, with $\zeta=1.8$. The underlying physical explanation for these powerlaws is left for future study.


{\it Conclusion:} 
Dirac and Weyl systems are shown to constitute a remarkably rich playground for Kondo physics, potentially exhibiting essentially the full range of single-impurity effects. In particular, the unconventional evolution of the Kondo temperature on varying chemical potential, and the unusual role of PH asymmetry, are `smoking gun' experimental signatures. 

We anticipate further interesting physics in topological Kondo lattice systems, and those with several magnetic impurities where pseudogap Kondo, RKKY, two-impurity Kondo, and disorder compete.


\acknowledgments
\emph{Acknowledgments:} 
The authors acknowledge former collaborations with M. Vojta, S. Florens, D, Schuricht, and R. Bulla. 
This work is part of the D-ITP consortium, a program of the Netherlands Organisation for Scientific Research (NWO) that is funded by the Dutch Ministry of Education, Culture and Science
(OCW).



\begin{thebibliography}{48}
\bibitem [{not({\natexlab{b}})}]{note:perts}%
  \BibitemOpen
  \bibinfo {note} {$\hat{\tau}_y\otimes\hat{\sigma}_0$ and
  $\hat{\tau}_z\otimes\vec{b}\cdot\hat{\vec{\sigma}}$ break both TRS and IS,
  leaving the Dirac nodes degenerate. The resulting impurity physics is
  equivalent to the TRS and IS invariant case.}\BibitemShut {Stop}%
\end{thebibliography}

%


\end{document}